\documentclass[preprint,aps]{revtex4}
\usepackage{amsmath}
\usepackage{amssymb}
\usepackage{axodraw}
\usepackage{graphicx}

\begin{document}

\title{Calculation  of pure annihilation type decay $B^+  \to D_s^+ \phi$
}

\author{Cai-Dian L\"u\footnote{e-mail: lucd@ihep.ac.cn}  }

\affiliation{ CCAST (World Laboratory), P.O. Box 8730,  Beijing 100080, China,\\
                           and\\
    Institute of High Energy Physics, CAS, P.O.Box 918(4)
 Beijing 100039, China\footnote{Mailing address.}
}

    \date{\today }

        \vfill


\begin{abstract}
The rare decay
 $B^+ \to D_s^+ \phi$ can occur only
via annihilation type diagrams in the standard model.
We calculate this decay
in  perturbative QCD approach with Sudakov resummation.
We found that the branching ratio of  $B^+  \to D_s^+ \phi$
 is of order $10^{-7}$
which may be  measured in the near future by KEK and SLAC $B$
factories.
  The small branching ratio predicted in standard model makes this channel
  sensitive to new physics contributions.
\end{abstract}

  \maketitle

\section{Introduction}

The rare $B$ decays are useful for the test of standard model (SM).
They are sensitive to new physics contributions, since their branching ratios in
SM are small. Some of them have already been measured  by CLEO and
 $B$ factories in KEK and SLAC.   Most of them are still on the
 way of study by both experimental and theoretical studies.
 Among them, the inclusive or semi-inclusive decays are clean in theory,
 but with more uncertainty in experimental study.
 On the other hand, the exclusive decays are difficult for precise theoretical
 prediction but easier for experimental measurement.
 The study of exclusive rare B decays require the knowledge of
 hadronization, which is non-perturbative.
The generalized factorization approach has been applied to the
theoretical treatment of non-leptonic  $B$ decays for some years \cite{bsw}.
It is a great success in explaining many decay branching ratios
\cite{akl1,cheng}.
The factorization approach is a rather simple method.
Some efforts have been made to improve their theoretical application
\cite{bbns} and to understand the reason why the factorization approach
has gone well \cite{Keum:2000,Lu:2000em}.
One of these method is the perturbative QCD approach (PQCD),
where we can calculate the annihilation diagrams as well as the factorizable
and non-factorizable diagrams.

The rare decay $B^+  \to D_s^+ \phi$
is pure annihilation type decay.
The four valence quarks in the final states  $D_s$ and $\phi$ are
different from the ones in B meson, i.e. there is no spectator
quark in this decay.
In the usual factorization approach, this decay picture is described as
$\bar b$ and $u$ quark in
$B$ meson annihilating into vacuum and the $D_s$ and $\phi$ meson produced
from vacuum then afterwards.
To calculate this decay in the factorization approach, one needs the
$D_s \to \phi$ form factor at very large time like momentum transfer
${\cal O} (M_B)$.
However the form factor at such a large momentum transfer is not known
in factorization  approach.
This makes the factorization approach calculation of these decays
unreliable.

In this paper, we will try to use the PQCD approach to calculate
this decay. The $W$ boson exchange causes the four quark operator transition
 $\bar b u \to \bar{s}c$, the additional $\bar{s}s$ quarks included in
$D_s \phi$ are produced from a gluon. This gluon  attaches to any one
of the quarks participating in $W$ boson exchange. This is shown
in Figure \ref{fig:diagrams2}. In the rest frame of $B$ meson,
both $s$ and $\bar{s}$ quarks included in $D_s \phi$ have
$\mathcal{O}(M_B)$ momenta, and the gluon producing them also has
momentum
$q^2 \sim \mathcal{O}(M_B^2)$. This is a hard gluon. One can
perturbatively treat the process where the four quark operator
 exchanges a hard gluon with $s \bar s$ quark pair.
It is just the picture of perturbative QCD approach
\cite{Keum:2000,Lu:2000em}. Furthermore, the final state of this
decay is isospin singlet. It is proportional to the $V_{ub}$
transition. Any $V_{cb}$ transition can not contribute to it.
Therefore there will not be any dominant soft final state
interaction contribution. Unlike the $B \to KK$ decays (which may
have large contribution from final state interaction contribution)
\cite{bkk}, the decay $B\to D_s \phi$ is a very clean channel for
a test of annihilation type contribution.

 In the next section, we
will show the framework of PQCD briefly. In section
\ref{sc:formula}, we give the analytic formulas for the decay
amplitude of $B^+  \to D_s^+ \phi$ decays. In section
\ref{sc:neval}, we give the numerical results of
 branching ratio from the
analytic formulas and discuss the theoretical errors.
Finally, we conclude this study in section \ref{sc:concl}.

\section{Framework}\label{sc:fm}

 PQCD approach has been developed and applied in  non-leptonic $B$
meson decays
\cite{Chang:1997dw,Yeh:1997rq,Keum:2000,Lu:2000em,bkk,Li:1999kn,pir}
for some time. In this approach, the decay amplitude is described
by three scale dynamics, soft($\Phi$), hard($H$), and harder($C$)
dynamics. It is conceptually written as the convolution,
\begin{equation}
 \mbox{Amplitude}
\sim \int\!\! d^4k_1 d^4k_2 d^4k_3\ \mathrm{Tr} \bigl[ C(t)
\Phi_B(k_1)  \Phi_{D_s}(k_2) \Phi_\phi(k_3) H(k_1,k_2,k_3, t)
\bigr], \label{eq:convolution1}
\end{equation}
where $k_i$'s are momenta of light quarks included in each mesons, and
$\mathrm{Tr}$ denotes the trace over Dirac and color indices.
$C(t)$ is Wilson coefficient of the four quark operator with the QCD radiative
corrections.
 $C(t)$ includes the harder dynamics
at larger scale than $M_B$ scale and describes the evolution of
local $4$-Fermi operators from $M_W$, $W$ boson mass, down to
$t\sim\mathcal{O}(M_B)$ scale, which results in large logarithms
$\ln (M_W/t)$. $H$ describes the four quark operator and the
 quark pair from sea connected by
 a hard gluon whose scale is at the order
of $M_B$, and includes the $\mathcal{O}(M_B)$ hard dynamics.
Therefore, this hard part $H$ can be perturbatively calculated.
$t$ is chosen as the largest energy scale in $H$, in order to
lower the $\alpha_s^2$ corrections to hard part $H$. $\Phi_M$ is
the wave function which describes hadronization of the quark and
anti-quark into the meson $M$. While  $H$ depends on the processes
considered, $\Phi_M$ is  independent on the specific processes.
Determining $\Phi_M$ in some other decays, we can make
quantitative predictions here.

We consider the $B$ meson at rest for simplicity. It is convenient
to use light-cone coordinate $(p^+, p^-, {\bf p}_T)$ to describe
the meson's momenta, where $p^\pm = (p^0 \pm p^3)/\sqrt{2}$ and
${\bf p}_T = (p^1, p^2)$. On this coordinate we can take the $B$,
$D_s$, and $\phi$ mesons momenta as $P_1 = \frac{M_B}{\sqrt{2}}
(1,1,{\bf 0}_T)$, $P_2 = \frac{M_B}{\sqrt{2}} (1,r^2,{\bf 0}_T)$,
and $P_3 = \frac{M_B}{\sqrt{2}} (0,1-r^2,{\bf 0}_T)$,
respectively, where $r = M_{D_s}/M_B$ and we neglect the square
terms of $\phi$ meson's mass $M_\phi^2$. Putting the light
spectator quark momenta for $B$, $D_s$ and $\phi$ mesons as $k_1$,
$k_2$, and $k_3$, respectively, we can choose
$k_1=(0,x_1P_1^-,{\bf k}_{1T})$, $k_2 = (x_2 P_2^+,0,{\bf
k}_{2T})$ and $k_3 = (0, x_3 P_3^-,{\bf k}_{3T})$. Then,
integration over $k_2^-$, $k_3^+$ and $k_1^-$ in
eq.(\ref{eq:convolution1}) leads to
\begin{multline}
 \mbox{Amplitude}
\sim \int\!\!
d x_1 d x_2 d x_3
b_1 d b_1 b_2 d b_2 b_3 d b_3 \\
\mathrm{Tr} \bigl[ C(t) \Phi_B(x_1,b_1)  \Phi_{D_s}(x_2,b_2)
\Phi_\phi(x_3, b_3) H(x_i, b_i, t) e^{-S(t)} \bigr],
\label{eq:convolution2}
\end{multline}
where $b_i$ is the conjugate space coordinate of $k_{iT}$.
 The last term, $e^{-S}$, contains two kinds of logarithms.
One of the large logarithms is due to the renormalization of
ultra-violet divergence $\ln tb$, which describes the QCD running between scale $t$
and $1/b$. The other is from double logarithm due to soft
gluon corrections.
This double logarithm called Sudakov form factor suppresses the soft dynamics effectively \cite{soft}.
Thus it makes perturbative calculation of the hard part $H$ applicable
at intermediate scale, i.e., $M_B$ scale.
We calculate the $H$ for $B^+  \to D_s^+ \phi$ decay in the first order in
$\alpha_s$ expansion and give the convoluted amplitudes in next section.

In order to calculate analytic formulas of the decay amplitude, we
use the wave functions $\Phi_{M,\alpha\beta}$ decomposed in terms
of spin structure.   As a heavy meson, $B$  meson wave function is
not well defined. It is also pointed out by  the recent discussion
of $B$ meson wave function \cite{qiao} that, there is no
constraint on $B$ meson wave function, if three-parton wave
functions are considered. To be consistent with previous
calculations \cite{Keum:2000,Lu:2000em,pir}, we follow the same
argument that    the structure
$(\gamma^\mu\gamma_5)_{\alpha\beta}$ and $\gamma_{5\alpha\beta}$
components make the dominant contribution in $B$ meson wave
function. Then, $\Phi_{M,\alpha\beta}$ is written by
\begin{equation}
 \Phi_{M,\alpha\beta} = \frac{i}{\sqrt{2N_c}}
\left\{
(\not \! P_M \gamma_5)_{\alpha\beta} \phi_M^A
+ \gamma_{5\alpha\beta} \phi_M^P
\right\},
\end{equation}
where $N_c = 3$ is color's degree of freedom,
$P_M$ is the corresponding meson's momentum, and
$\phi_M^{A,P}$ are Lorentz scalar wave functions.
As heavy quark effective theory leads to
$\phi_B^P \simeq M_B \phi_B^A$, then $B$ meson's wave function can be
expressed by
\begin{equation}
 \Phi_{B,\alpha\beta}(x,b) = \frac{i}{\sqrt{2N_c}}
\left[ \not \! P_1  + M_B \right]\gamma_{5\alpha\beta}
\phi_B(x,b).
\end{equation}

According to Ref.\cite{Ball:1998je}, a pseudo-scalar meson moving fast
is parameterized by Lorentz scalar wave functions, $\phi$, $\phi_p$,
and $\phi_\sigma$ as
\begin{gather}
\langle D_s^-(P)|{\bar s}(z)\gamma_\mu \gamma_5 c(0)| 0 \rangle \simeq
- i f_{D_s} P_\mu\int_0^1\!\!\! dx\ e^{ix P z}\phi (x),
\label{pv} \\
\langle D_s^-(P)|{\bar s}(z)\gamma_5 c(0)|0 \rangle =
-if_{D_s} m_{0D_s} \int_0^1\!\!\! dx\ e^{ix P z}\phi_p(x),
\label{ps} \\
\langle D_s^-(P)|{\bar s}(z)\gamma_5 \sigma_{\mu\nu} c(0)|0 \rangle =
\frac{i}{6}f_{D_s} m_{0D_s} \left(1-\frac{M_{D_s}^2}{m_{0D_s}^2} \right)
(P_\mu z_\nu-P_\nu z_\mu)
\int_0^1\!\!\! dx\ e^{ix P z}\phi_\sigma(x),
\label{pt}
\end{gather}
where $m_{0D_s} = M_{D_s}^2/(m_c+m_s)$. We ignore the difference
between $c$ quark's mass and $D_s$ meson's mass in the
perturbative calculation. This means,  $ M_{D_s} = m_{0D_s}$, In
this approximation, the contributions of eq.(\ref{pt}) are
negligible. With the equation of motion    eq.(\ref{pv}),
eq.(\ref{ps}),  lead to
\begin{equation}
\phi_p(x) = \phi(x) +
{\cal O}\left( \frac{\bar{\Lambda}}{M_{D_s}}\right).
\end{equation}
Therefore
the $D_s$ meson's wave function can be expressed by one
Lorentz scalar wave function,
\begin{equation}
 \Phi_{D_s,\alpha\beta}(x,b) = \frac{i}{\sqrt{2N_c}}
\left[
(\gamma_5 \not \! P_2 )_{\alpha\beta}
+ M_{D_s} \gamma_{5\alpha\beta}
\right]   \Phi_{D_s}(x,b).
\end{equation}
The wave function $\Phi_M$ for $M = B, D_s$ meson is normalized by its
decay constant $f_M$
\begin{equation}
 \int_0^1 \!\! dx\  \Phi_M (x, b=0)
= \frac{f_M}{2\sqrt{2N_c}}.
\label{eq:normalization}
\end{equation}

In contrast to the $B$ and $D_s$ meson,
for the $\phi$ meson, being light,
the $\sigma^{\mu\nu}_{\alpha\beta}$ component remains.
In $B^+ \to D_s^+ \phi$ decay, the $\phi$ meson is longitudinally
polarized.
Then, $\phi$ meson's wave function is parameterized by  three
Lorentz structures
\begin{align}
 \frac{ M_\phi \not \epsilon}{\sqrt{2N_c}}
  \Phi_\phi(x_3),     ~~
  \frac{\not \epsilon \not \! P_3}{\sqrt{2N_c}}  \Phi_\phi^t(x_3),
  ~~
     \frac{ M_\phi }{\sqrt{2N_c}}
  \Phi_\phi^s(x_3)   .
\end{align}
In the numerical analysis we will use $\Phi_\phi$, $ \Phi_\phi^t$
and $ \Phi_\phi^s$ which were calculated from QCD sum rule
\cite{sumrule}. They will be shown in section \ref{sc:neval}.

\section{Perturbative Calculations}\label{sc:formula}

         \begin{figure}[htbp]
       \scalebox{0.7}{
         \begin{picture}(130,110)(0,-10)
            \ArrowLine(50,50)(13,63)
            \ArrowLine(13,37)(50,50)
            \Line(55,50)(72.5,32.5)
            \ArrowLine(72.5,32.5)(95,10)
            \Line(55,50)(72.5,67.5)
            \ArrowLine(95,90)(72.5,67.5)
            \ArrowLine(90,50)(115,75)
            \ArrowLine(115,25)(90,50)
            \Gluon(72.5,67.5)(90,50){3}{4}
            \Vertex(90,50){1.5} \Vertex(72.5,67.5){1.5}
            \put(0,47){$B^+$}
            \put(108,7){$D_s^+$}
            \put(108,89){$\phi$}
                \put(20,70){\small{$\bar{b}$}}
                     \put(18,28){\small{$u$}}
                  \put(75,14){\small{$c$}}
                         \put(77,84){\small{$\bar s$}}
            \put(50,-5){(a)}
         \end{picture}
       }
    \scalebox{0.7}{
      \begin{picture}(130,110)(0,-10)
         \ArrowLine(50,50)(13,63)
         \ArrowLine(13,37)(50,50)
         \Line(55,50)(72.5,32.5)
         \ArrowLine(72.5,32.5)(95,10)
         \Line(55,50)(72.5,67.5)
         \ArrowLine(95,90)(72.5,67.5)
         \ArrowLine(90,50)(115,75)
         \ArrowLine(115,25)(90,50)
         \Gluon(72.5,32.5)(90,50){3}{4}
         \Vertex(90,50){1.5} \Vertex(72.5,32.5){1.5}
         \put(0,47){$B^+$}
         \put(108,7){$D_s^+$}
         \put(108,89){$\phi$}
         \put(50,-5){(b)}
      \end{picture}
    }
    \scalebox{0.7}{
      \begin{picture}(120,130)(0,-20)
            \ArrowLine(50,50)(13,63)
            \ArrowLine(13,37)(26,41.8)      \ArrowLine(26,41.8)(50,50)
            \Line(55,50)(72.5,32.5)
            \ArrowLine(72.5,32.5)(95,10)
            \Line(55,50)(72.5,67.5)
            \ArrowLine(95,90)(72.5,67.5)
            \ArrowLine(90,50)(115,75)
            \ArrowLine(115,25)(90,50)
                       \GlueArc(46,160)(120,261,292){4}{9}
            \Vertex(90,50){1.5} \Vertex(26,41.8){1.5}
         \put(0,45){$B^+$}
         \put(105,3){$D_s^+$}
         \put(105,87){$\phi$}
         \put(20,70){\small{$\bar{b}$}}
              \put(18,28){\small{$u$}}
           \put(75,14){\small{$c$}}
                   \put(77,84){\small{$\bar s$}}
         \put(50,-20){(c)}
      \end{picture}
    }
     \scalebox{0.7}{
       \begin{picture}(120,130)(0,-20)
             \ArrowLine(50,50)(26.5,57.9)    \ArrowLine(26.5,57.9)(13,63)
             \ArrowLine(13,37)(50,50)
             \Line(55,50)(72.5,32.5)
             \ArrowLine(72.5,32.5)(95,10)
             \Line(55,50)(72.5,67.5)
             \ArrowLine(95,90)(72.5,67.5)
             \ArrowLine(90,50)(115,75)
             \ArrowLine(115,25)(90,50)
                        \GlueArc(46.2,-60.5)(120,68.5,99){4}{9}
             \Vertex(90,50){1.5} \Vertex(26.5,57.9){1.5}
          \put(0,45){$B^+$}
          \put(105,3){$D_s^+$}
          \put(105,87){$\phi$}
          \put(20,70){\small{$\bar{b}$}}
               \put(18,28){\small{$u$}}
            \put(75,14){\small{$c$}}
                    \put(77,84){\small{$\bar s$}}
          \put(50,-20){(d)}
       \end{picture}
     }
    \caption{Diagrams for $B^+ \to D_s^+ \phi$ decay. The factorizable
    diagrams (a),(b) contribute to $F_a $, and non-factorizable (c),
    (d) do to $M_a $.}
    \label{fig:diagrams2}
   \end{figure}
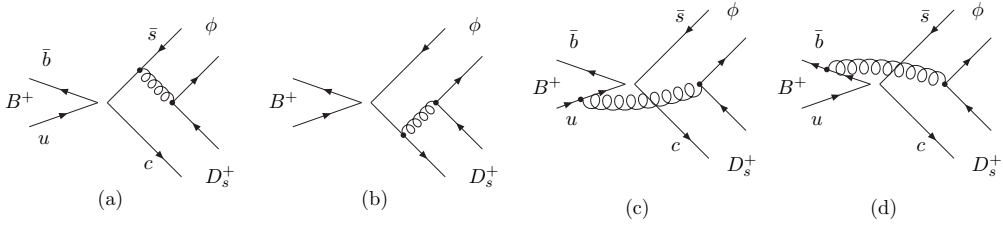

  The effective Hamiltonian related to
  $B^+ \to D_s^+ \phi$ decay is given as  \cite{Buchalla:1996vs}
  \begin{gather}
   H_\mathrm{eff} = \frac{G_F}{\sqrt{2}} V_{ub}^*V_{cs} \left[
  C_1(\mu) O_1(\mu) + C_2(\mu) O_2(\mu) \right] \\
    O_1 = (\bar{b} \gamma_\mu P_L s) (\bar{c} \gamma^\mu P_L u) , \quad
   O_2 = (\bar{b} \gamma_\mu P_L u)  (\bar{c} \gamma^\mu P_L s) .
  \end{gather}

where $C_{1,2}(\mu)$ are Wilson coefficients at renormalization scale
$\mu$. The projection operator is defined as
$P_L=1-\gamma_5$.
The lowest order diagrams contributing to $B^+ \to D_s^+ \phi$ are
drawn in Fig.\ref{fig:diagrams2} according to this effective
Hamiltonian.
As stated above, $B^+ \to D_s^+ \phi$ decay only has  annihilation
diagrams.

We get the following analytic formulas by calculating the hard part $H$
at first order in $\alpha_s$.
Together with the meson wave functions,
the amplitude for the factorizable annihilation diagram
in Fig.\ref{fig:diagrams2}(a) and (b)
results in,
\begin{multline}
F_a = 16\pi C_F f_B M_B^2 \int_0^1\!\!\! dx_2 dx_3
 \int_0^\infty\!\!\!\!\!  b_2 db_2\, b_3 db_3\  \Phi_{D_s}(x_2,b_2)
\times \Bigl[ \bigl\{
  x_3 \Phi_\phi(x_3,b_3)          \\
+ r \left(  2  x_3-1 \right) r_\phi \Phi_\phi^t(x_3,b_3) \
+ r   (1 + 2 x_3 ) r_\phi \Phi_\phi^s(x_3,b_3)
\bigr\} E_{f} (t_a^1) h_a(x_2,x_3,b_2,b_3) \\
- \bigl\{
 x_2\Phi_\phi (x_3,b_3)
+ 2 r (1 +x_2) r_\phi \Phi_\phi^s(x_3,b_3)
\bigr\}
E_{f} (t_a^2) h_a(x_3,x_2,b_3,b_2) \Bigr],
\label{eq:Fa}
\end{multline}
where $C_F = 4/3$ is the group factor of $\mathrm{SU}(3)_c$ gauge
group, and $r_\phi = m_{\phi}/M_B$.  The function $E_{f}$,
$t_a^{1,2}$, $h_a$ are given in the Appendix. Since we only
include twist 2 and twist 3 contributions in our PQCD approach,
all the $r^2$ and $r_\phi^2$ terms in the calculation are
neglected for consistence. The explicit form for the wave
functions, $\Phi_M$, is given in the next section.
 From eq.(\ref{eq:Fa}), one can see that the factorizable
 contribution $F_a$ is independent of   the $B$ meson wave
 function, but proportional to the $B$ meson decay constant $f_B$.

The amplitude for the non-factorizable annihilation diagram in
Fig.\ref{fig:diagrams2}(c) and (d) is given as
\begin{multline}
M_a =  \frac{1}{\sqrt{2N_c}} 64\pi C_F M_B^2
\int_0^1\!\!\! dx_1 dx_2 dx_3
 \int_0^\infty\!\!\!\!\! b_1 db_1\, b_2 db_2\
 \phi_B(x_1,b_1)  \Phi_{D_s}(x_2,b_2) \\
\times \Bigl[ \bigl\{ x_2 \Phi_\phi(x_3,b_2)
 + r \left(x_2-x_3\right) r_\phi \Phi_\phi^t(x_3,b_2) \\
 + r \left(x_2+ x_3\right) r_\phi \Phi_\phi^s(x_3,b_2)
\bigr\}
E_{m}(t_{m}^1) h_a^{(1)}(x_1,x_2,x_3,b_1,b_2) \\
- \bigl\{
    x_3 \Phi_\phi(x_3,b_2)
 - r \left( x_2 -x_3 \right) r_\phi \Phi_\phi^t(x_3,b_2) \\
 + r \left(2+x_2 +x_3 \right) r_\phi \Phi_\phi^s(x_3,b_2)
\bigr\}E_{m}(t_{m}^2) h_a^{(2)}(x_1,x_2,x_3,b_1,b_2) \Bigr].
\label{eq:Ma2}
\end{multline}
   Unlike the factorizable
contribution $F_a$, the non-factorizable annihilation diagram
involve all three meson wave functions.

Thus, the total decay amplitude $A$ and decay width $\Gamma$ for
$B^+ \to D_s^+ \phi$ decay are given as
\begin{gather}
  A =  F_a + M_a,
\label{eq:chrg_amp} \\
 \Gamma(B^+ \to D_s^+ \phi) = \frac{G_F^2 M_B^3}{128\pi}
|V_{ub}^*V_{cs} A|^2 ,\label{eq:chrg_width}
\end{gather}
 where the overall factor is included in the decay width
 with the kinematics factor.

The decay amplitude for CP conjugated mode, $B^- \to D_s^- \phi$,
is the same expression as $B^+ \to D_s^+ \phi$,    just replacing
$V_{ub}^*V_{cs}$ with $V_{ub}V_{cs}^*$.
Since there is only one kind of CKM phase involved in the decay, there is
no CP violation in the standard model for this channel.
 We thus have $Br(B^+\to D_s^+\phi)=Br(B^-\to D_s^- \phi)$.

\section{Numerical Results}\label{sc:neval}

In this section we show numerical results obtained from the previous
formulas.
At the beginning, we give the branching ratios predicted from the same
parameters and wave functions that are adopted in   other works.
Secondly, we discuss the theoretical errors due to uncertainty of some
parameters.

For the $B$  meson's wave function, there is a sharp peak at the small $x$
region, we use
\begin{equation}
\Phi_B(x,b) = N_B x^2(1-x)^2 \exp \left[
-\frac{M_B^2\ x^2}{2 \omega_b^2} -\frac{1}{2} (\omega_b b)^2
\right],
\end{equation}
which is adopted in ref. \cite{Keum:2000,Lu:2000em,pir}. This
choice of $B$ meson's wave function is almost a best  fit from the
$B\to K\pi$, $\pi \pi$, $\pi\rho$ and $\pi\omega$ decays. For the
$D_s$ meson's wave function, we assume the  form as the following,
leaving $a_{D_s}$ a free parameter
\begin{equation}
 \Phi_{D_s}(x,b) = \frac{3}{\sqrt{2 N_c}} f_{D_s}
x(1-x)\{ 1 + a_{D_s} (1 -2x) \} \exp \left[  -\frac{1}{2}
(\omega_D b)^2\right].
\end{equation}
This is a rather flat distribution function. Since $c$ quark is
 heavier than $s$ quark, this function is peaked at $c$ quark
side, i.e. small $x$ region. The wave functions  of the $\phi$
meson are derived by QCD sum rules \cite{sumrule}
\begin{eqnarray}
\Phi_\phi(x) &=& \frac{f_\phi}{2\sqrt{2 N_c}} 6 x(1-x)  ,
 \\
\Phi_\phi^t(x) &=& \frac{f_\phi^T}{2\sqrt{2 N_c}}
\left\{  3\xi^2 + 0.21 \left(3 - 30\xi^2 +35\xi^4\right)
 +0.69  \left (1+\xi  \ln \frac{x}{1-x} \right) \right\},
  \\
\Phi_\phi^s(x) &=& \frac{f_\phi^T}{4\sqrt{2 N_c}}
\left\{ 3\xi\left (4.5 -11.2x +11.2~x^2\right) +1.38 \ln \frac{x}{1-x}  \right\},
\end{eqnarray}
where $\xi =1- 2 x $.
In addition, we use the following input parameters:
\begin{gather}
 M_B = 5.279 \mbox{ GeV},\  M_{D_s} = 1.969 \mbox{ GeV},
m_{\phi} = 1.02 \mbox{ GeV},
\label{eq:parm1} \\
f_B = 190 \mbox{ MeV},\  f_\phi = 237 \mbox{ MeV},\ f_\phi^T = 220 \mbox{ MeV},\
f_{D_s} = 241 \mbox{ MeV}, \\
\omega_b = 0.4 \mbox{ GeV},\  a_{D_s} = 0.3,\ \omega_D=0.2GeV.
\label{eq:shapewv}
\end{gather}
With these values  and eq.(\ref{eq:normalization}) we get the
normalization factor $N_B = 91.745$ GeV. Using the above fixed
parameters,  we find that  the factorizable annihilation diagram
contribution is dominant over  the non-factorizable contribution.
 The reason is that the Wilson coefficient in non-factorizable
 contribution $M_a$ is $C_1(t)$, which is smaller
than the one in factorizable contribution $F_a $, $a_1=C_1/3
+C_2$. Although the real part of $M_a$ is negligible, the
imaginary part of $M_a$ is comparable with the imaginary part of
$F_a$, it is about 30\% of the real part of $F_a$.

The propagators of inner quark and gluon in Figure 1 are usually
proportional to  $1/x_i$. One may suspect that these amplitudes
are enhanced by the endpoint singularity around $x_i \sim 0$. This
can be explicitly found in eq.(\ref{eq:propagator1},
\ref{eq:propagator2}), where the Bessel function $\mathrm{Y}_0$
diverges at $x_i \sim 0$ or $1$. However this is not the truth in
our calculation. First we introduce the transverse momentum of
quark, such that the propagators become $1/(x_i x_j +k_T^2)$.
There is no divergent at endpoint region. Secondly, the Sudakov
form factor $\mathrm{Exp}[-S]$ suppresses the region of small
$k^2_T$. Therefore there is no singularity in our calculation. We
also include the threshold resummation in our calculation of
factorizable diagrams, which further suppress the endpoint region
contribution \cite{thre}. The dominant contribution is not from
the endpoint region of the wave function. As a prove, in our
numerical calculations, for example, an expectation value of
$\alpha_s$ in the integration for $F_a$ and $M_a$ results in
$\langle \alpha_s/\pi \rangle \simeq 0.1$. Therefore, the
perturbative calculations are self-consistent.


Now we can calculate the branching ratio according to eqs.
(\ref{eq:chrg_amp},
\ref{eq:chrg_width}).
Here we use CKM matrix elements \cite{pdg}
\begin{gather}
  |V_{ub} |=0.0036\pm 0.0010, \
  |V_{cs}|=0.9891\pm 0.016,
\label{eq:KMmatrix}
\end{gather}
and the life time for $B^\pm$ meson is
 $\tau_{B^\pm}=1.65\times 10^{-12}\mbox{ s}$.
The predicted branching ratio is
\begin{gather}
 \mathrm{Br}(B^+ \to D_s^+ \phi) = 3.0 \times 10^{-7}.
\label{eq:br2}
\end{gather}
This is still far from  the current experimental upper limit
\cite{pdg}
\begin{gather}
 \mathrm{Br}(B^+ \to D_s^+ \phi ) < 3.2 \times 10^{-4}.
\end{gather}

The branching ratios obtained from the analytic formulas may be sensitive
to various parameters,
such as  parameters in
eqs.(\ref{eq:shapewv}).
Uncertainty of the predictions on PQCD is mainly due to the meson wave
functions.
Therefore it is important to give the limits of the branching ratio
when we choose the parameters to appropriate extent.
The appropriate extent of $\omega_b$ can be obtained from calculation
 of semi-leptonic decays \cite{kurimoto} and other $B\to \pi \pi $, $B\to K\pi$
 and $B\to \rho\pi$, $\omega\pi$ decays
 \cite{Keum:2000,Lu:2000em,pir},
\begin{equation}
 0.35 \mbox{ GeV} \leq \omega_b \leq 0.45 \mbox{ GeV}.
\end{equation}
The change of value of $\omega_b$ will not alter the result of
$F_a$, which is independent of $B$ meson wave function, but will
affect the value of $M_a$. We did not find any strict constraints
for the $D_s$ meson wave function in the literature. In fact, a
future study of $B\to D_s\pi$ will do this job. At present,
$a_{D_s}$ in $D_s$ meson wave function is a free parameter, and we
take $0 \leq a_{D_s} \leq 1$. Here we check the sensitivity of our
predictions on $\omega_b$ and $a_{D_s}$ within the ranges stated
above. The branching ratios normalized by the decay constants and
the CKM matrix elements can result in
\begin{gather}
\mathrm{Br}(B^+ \to D_s^+ \phi  ) = (3.0 ^{+2.4}_{-1.0}) \times
10^{-7} \left( \frac{f_B\ f_{D_s}}{190\mbox{ MeV}\cdot 241\mbox{
MeV}} \right)^2 \left( \frac{|V_{ub}^*\ V_{cs}|} { 0.0036 \cdot
0.9891} \right)^2 .
\end{gather}
Considering the  uncertainty of $f_B$, $f_{D_s}$ and $|V_{ub}^*
V_{cs}|$ etc., the branching ratio of
 $B^+ \to D_s^+ \phi$ decay is at the order of $10^{-7}$.
 This may be
measured by the current $B$ factory experiments in KEK and SLAC.


\section{Conclusion}\label{sc:concl}

In two-body hadronic $B$ meson decays, the final state mesons are
moving very fast, since each of them carry more than 2 GeV energy.
There is not enough time for them to exchange soft gluons. The
soft final state interaction is not important in the two-body $B$
decays. This is consistent with the argument based on
color-transparency \cite{bjo}. The PQCD with Sudakov form factor
is a self-consistent approach to describe the two-body $B$ meson
decays. Although the  annihilation diagrams are suppressed
comparing to other spectator diagrams, but their contributions are
not negligible in PQCD approach \cite{Keum:2000,Lu:2000em}.

In this paper, we calculate the $B^+ \to D_s^+ \phi$ decay in the
PQCD approach. Since neither of the bottom quark nor the up quark
in the initial B meson appeared in the final mesons, this process
 occurs purely via annihilation type diagrams. It is a charm quark (not an anti-charm quark)
 in the final states, therefore the usual $V_{cb}$ transition does not
 contribute to this process. The final states are isospin singlet.
  There should be no dominant final state interactions through
 other channels contribute.
 From our PQCD study, the branching ratio of  $B^+ \to D_s^+ \phi$
decay is still sizable with a branching ratio around $10^{-7}$,
which  may be measured  in the current running $B$ factories
Belle, BABAR or in LHC-B in the future. This may be one of the
channels to be measured in $B$ decays via annihilation type
diagram. Whether the PQCD predicted branching ratio is good enough
to account for the $B^+\to D_s^+ \phi$ decay will soon be tested
in the current or future experiments.

 The small branching ratio
(comparing to the already measured other B decays) predicted in
the SM, makes this channel sensitive to any new physics
contributions. Since the CP asymmetry predicted for this channel
in SM is zero, any non-zero measurement of CP asymmetry will be a
definite signal of new physics. We also notice that the
supersymmetric contribution will not enhance the decay branching
ratio significantly, but it may contribute to a non zero CP
asymmetry in this channel, since the supersymmetry couplings can
introduce new phases.

\section*{Acknowledgments}

We thank our PQCD group members: Y.Y. Keum, E. Kou, T. Kurimoto,
H.-n. Li,  A.I. Sanda and  M.Z. Yang   for fruitful discussions.
The work is partly supported by the Grand-in Aid for Special
Project Research (Physics of CP violation) of Japan.

\begin{appendix}

\section{Some functions}

The definitions of some functions used in the text are presented
in this appendix.
In the numerical analysis we use one loop expression for strong coupling
constant,
\begin{equation}
 \alpha_s (\mu) = \frac{4 \pi}{\beta_0 \log (\mu^2 / \Lambda^2)},
\label{eq:alphas}
\end{equation}
where $\beta_0 = (33-2n_f)/3$ and $n_f$ is number of active quark flavor at
appropriate scale. $\Lambda$ is QCD scale, which we use as $250$ MeV at
$n_f=4$.
We also use leading logarithms expressions for Wilson coefficients
$C_{1,2}$ presented in ref.\cite{Buchalla:1996vs}.
Then, we put $m_t = 170$ GeV, $m_W = 80.2$ GeV, $m_b = 4.8$ GeV,
and $m_c = 1.3$ GeV in the Wilson coefficients calculation.

The function $E_f$ and $E_m$
are defined as
\begin{eqnarray}
 E_{f}(t) &=& \left[ C_1(t)/3 + C_2(t) \right] \alpha_s(t)\, e^{-S_D(t)-S_\phi(t)}, \\
 E_{m}(t) &=& C_1(t) \alpha_s(t)\, e^{-S_B(t)-S_D(t)-S_\phi(t)}.
\end{eqnarray}
The above $S_{B, D, \phi}$ are defined as
\begin{eqnarray}
S_B(t) &=& s(x_1P_1^+,b_1) +
2 \int_{1/b_1}^t \frac{d\mu'}{\mu'} \gamma_q(\mu'), \\
S_D(t) &=& s(x_2P_2^+,b_3) +
2 \int_{1/b_2}^t \frac{d\mu'}{\mu'} \gamma_q(\mu'), \\
S_\phi(t)& =& s(x_3P_3^+,b_3) + s((1-x_3)P_3^+,b_3) +
2 \int_{1/b_3}^t \frac{d\mu'}{\mu'} \gamma_q(\mu'),
\end{eqnarray}
where the last terms of the above formulas
are logarithms from   the renormalization of ultra-violet divergence.
   The term
$s(Q,b)$, the so-called Sudakov factor,
  result from summing  up double logarithms
       caused by collinear divergence and soft divergence.
The expression  is given as
\cite{Li:1999kn}
\begin{eqnarray}
  s(Q,b) &=& \int_{1/b}^Q \!\! \frac{d\mu'}{\mu'} \left[
 \left\{ \frac{2}{3}(2 \gamma_E - 1 - \log 2) + C_F \log \frac{Q}{\mu'}
 \right\} \frac{\alpha_s(\mu')}{\pi} \right. \nonumber \\
& &  \left.+ \left\{ \frac{67}{9} - \frac{\pi^2}{3} - \frac{10}{27} n_f
 + \frac{2}{3} \beta_0 \log \frac{\gamma_E}{2} \right\}
 \left( \frac{\alpha_s(\mu')}{\pi} \right)^2 \log \frac{Q}{\mu'}
 \right],
 \label{eq:SudakovExpress}
\end{eqnarray}
 $\gamma_E=0.57722\cdots$ is Euler constant,
and $\gamma_q = \alpha_s/\pi$ is the quark anomalous dimension.

The $h$'s in the decay amplitudes are given by performing Fourier
transformation on the transverse momenta ${\bf k}_{iT}$ for propagators
of virtual quark and gluon in the hard part calculation, they result in
\begin{eqnarray}
 h_a(x_2,x_3,b_2,b_3) &=& \left( \pi i/2\right)^2
H_0^1(M_B\sqrt{ x_2 x_3} b_2) S_t(x_3) \label{eq:propagator1}\\
&&\times \left\{
H_0^1(M_B\sqrt{ x_3} b_2) J_0(M_B\sqrt{ x_3} b_3)
\theta(b_2 - b_3) + (b_2 \leftrightarrow b_3 ) \right\},    \nonumber
 \\
 h^{(j)}_a(x_1,x_2,x_3,b_1,b_2)& =&
   \left(
   \begin{matrix}
    \mathrm{K}_0(M_B  \sqrt{F_j} b_1), & \text{for}\quad F_j \ge 0 \\
    \frac{\pi i}{2} \mathrm{H}_0^{(1)}(M_B\sqrt{ -F_j}\ b_1), &
    \text{for}\quad F_j<0
   \end{matrix}\right)  \times     \label{eq:propagator2}
  \\
&& \left\{ \frac{\pi i}{2} \mathrm{H}_0^{(1)}(M_B\sqrt{ x_2 x_3}b_1)
 \mathrm{J}_0(M_B\sqrt{ x_2x_3}b_2) \theta(b_1-b_2)
+ (b_1 \leftrightarrow b_2) \right\},
          \nonumber
\end{eqnarray}
with the variables $F_1=x_2 (x_1-x_3)$, $F_2=x_2+ (1 -x_2)(x_1+
x_3)$. And $\mathrm{H}_0^{(1)}(z) = \mathrm{J}_0(z) + i\,
\mathrm{Y}_0(z)$.
 The threshold resummation form factor $S_t(x_i)$ is
adopted from ref.\cite{kurimoto}
\begin{equation}
S_t(x)=\frac{2^{1+2c} \Gamma (3/2+c)}{\sqrt{\pi} \Gamma(1+c)}
[x(1-x)]^c,
\end{equation}
where the parameter $c=0.3$. This function is normalized to unity.
The hard scale $t$'s in the amplitudes are taken as the largest
energy scale in  $H$ to diminish the higher order $\alpha_s^2$
corrections:
\begin{gather}
 t_a^1 = \mathrm{max}(M_B \sqrt{ x_3},1/b_2,1/b_3), \\
 t_a^2 = \mathrm{max}(M_B \sqrt{ x_2},1/b_2,1/b_3), \\
 t_{m}^1 = \mathrm{max}(M_B \sqrt{|F_1|}, M_B \sqrt{ x_2 x_3 },1/b_1,1/b_2),                 \\
  t_{m}^2 = \mathrm{max}(M_B  \sqrt{F_2},
 M_B \sqrt{ x_2 x_3 }, 1/b_1,1/b_2).
\end{gather}

\end{appendix}

\end{document}